# High-Gradient, Millimeter Wave Accelerating Structure

S.V. Kuzikov[1,2], A.A. Vikharev[1,3], N.Yu. Peskov[1]

[1]*Institute of Applied Physics,
46 Ulyanova Str., Nizhny Novgorod, 603950, Russia*
[2]*Omega-P Inc., New Haven, CT 06511*
[3]*Yale University, New Haven, CT 06520-8120*

The millimeter wave all-metallic accelerating structure, aimed to provide more than 100 MeV/m gradient and fed by feeding RF pulses of 20-30 ns duration, is proposed. The structure is based on a waveguide with small helical corrugation. Each section of 10-20 wavelengths long has big circular cross-section aperture comparable with wavelength. Because short wavelength structures are expected to be critical to wakefields excitation and emittance growth, we suggest to combine in one structure properties of a linear accelerator and a cooling damping ring simultaneously. It provides acceleration of straight on-axis beam as well as cooling of this beam due to the synchrotron radiation of particles in strong non-synchronous transverse fields. These properties are provided by specific slow eigen mode which consists of two partial waves, $TM_{01}$ and $TM_{11}$. Simulations show that shunt impedance can be as high as 100 MOhm/m. Results of the first low-power tests with 30 GHz accelerating section are analyzed.

Insert PSN Here

# 1. BASIC PRINCIPLES

Recently Tantawi et al. (2014) showed GV/m level accelerating gradient in 116 GHz structure fed by short rf pulse produced by FACET's electron bunch. This experiments inspires to invent mm-wave accelerating structures having new properties. In particular, a millimeter wave structure should be fed by short, high repetition rate rf pulses ($\tau$<20 ns), in order to avoid breakdown and excessive pulse heating according to scaling laws for surface electric $E_s$ and magnetic $H_s$ fields:

$$E_s^p \tau = const, \; p = 5-6.$$
$$H_s^2 \sqrt{\tau} = const.$$
(1)

A mm-wave structure should also have: wide (≥1 cm) channel in whole section without junctions for electron beam channeling and efficient pumping; high coupling coefficient of cells to avoid strong sensitivity to spread of geometric parameters; smooth transverse beam focusing due to the ponderomotive (Miller's) force is also desirable.

Next important property for a mm-wave structure follows from that emittances of bunches propagating in high-gradient accelerator have tendency to increase. Because loss factor for excitation of wakefields is scaled as ~$1/a^2$ ($a$ – structure's aperture), and it is expected that beam emittance grows up with wavelength even faster (due to wakefields excited by geometry mistakes), one should preserve bunch emittances and energy spread as well directly in the structure. An accelerating RF structure, which also plays a role of cooling RF undulator, can satisfy the mentioned requirements [1-3]. Such structure might be based on helical corrugated waveguide, where the operating normal mode consists of the $0^{th}$ spatial harmonic (with positive phase velocity), which is actually the accelerating mode, and the -$1^{st}$ harmonic (with negative phase velocity) which is responsible for transverse particle wiggling (Fig. 1). The transverse non-synchronous field components can provide emittance control and beam focusing due to ponderomotive force which depends on transverse electric field gradient. A new structure has smooth shape of the constant circular cross-section (no expansions or narrowings) and big aperture (no small irises). Perhaps, a new technology of the mass production, based on a "corkscrew" in a copper mandrel, seems also possible which allows avoiding junctions inside sections.

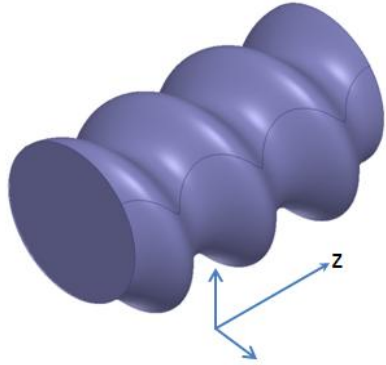

Figure 1: Shape of helical accelerating structure.

## 2. TM01-TM11 HELICAL ACCELERATING STRUCTURE

### 2.1. Optimization of Parameters

Insert PSN Here

For optimization we chose helical structure shape given in cylindrical system of coordinates ($r$, $z$, $\varphi$) by equation:

$$r(z,\varphi) = R + a \cdot \sin(\frac{2\pi \cdot z}{L} + m\varphi) \quad (2)$$

where $R$ – is an average waveguide radius, $a$ and $L$ – are amplitude and period of the corrugation. Period of the corrugation is close to $2\pi/h_{TM01}$, where $h_{TM01}$ – is a propagation constant of the partial TM$_{01}$ mode in smooth circular waveguide of the radius $R$. Such helical profile couples partial waves TM$_{01}$ and rotating on azimuth TM$_{11}$. Parameters of tthis structure was optimized to reach maximum of the accelerating field relative to maximum of the arisen surface field $E_a/E_s^{max}$, maximum of shunt impedance $R_{sh}$ was also necessary. Results of optimization for geometry at frequency 28.2 GHz are $E_a/E_s^{max} = 0.31$, $R_{sh} = 19$ MOhm/m for $R$=6.09 mm, $L$=8 mm, $a$=1.25 mm. For example, if the gradient is as high as 100 MV/m, then equivalent transverse magnetic field reaches 0.75 T so that the beam of 25 GeV energy has decay distance as long as 2800 m.

Dispersion relation for the optimized structure (Fig. 2) in a form of dependence of the eigen frequency $f$ on phase advance $h_0L$ ($h_0$ – is a Floquet propagation constant) is plotted in Fig. 3. The line 1 in Fig. 3 is actually light cone which corresponds to the condition of Cherenkov's synchronism. The operating mode corresponds the cross point of the curves 1 and 2. In this point the eigenmode has a property that phase velocities of accelerating component and all transverse components of electric or magnetic fields have opposite signs (Fig. 4).

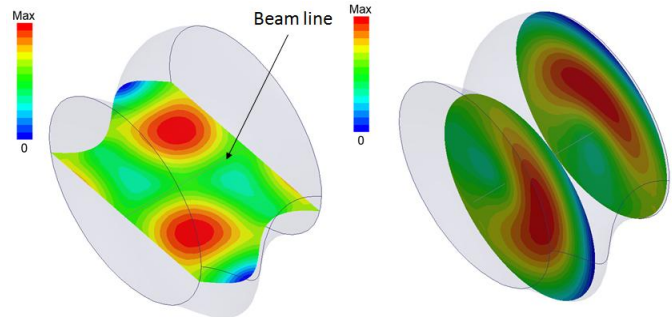

Figure 2: Eigenmode of TM01-TM11 helical accelerating structure: complex of electric field in middle plane (left) and in two sequent cross-sections (right).

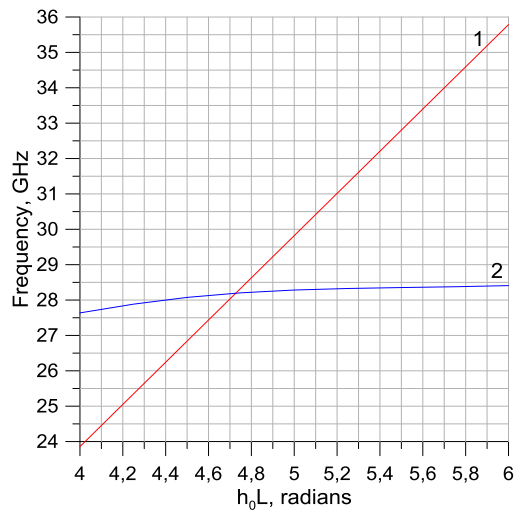

Figure 3: Frequency dispersion in helical waveguide: light cone line (1) and eigenmode dispersion curve (2).

Insert PSN Here

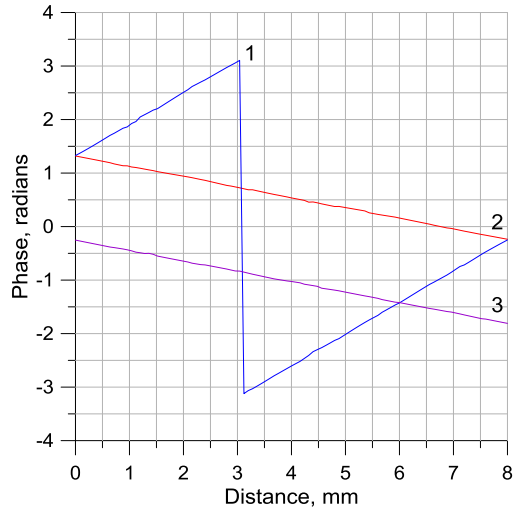

Figure 4: Phase diagram for operating mode: phase of accelerating field (1) and phases of x- electric field component (2) and y-component (3).

## 2.2. Low-Power Test

For test at low-power level we produced the helical structure for frequency 29.3 GHz (Fig. 5). Because we decided to feed our helical structure by $TM_{01}$ incident mode coming from smooth waveguide of radius equal to average radius of the structure, we considered two cases. In the first case we optimized structure's full length, in order to minimize reflection near the operating frequency (Fig. 6). In the second case we added two helical sections at both ends of the structure twisted relative to the main structure (Fig. 7). These sections correspond actually to Chebyshev's matching principle [4]. In both cases we used $TE_{11}$-$TM_{01}$ mode converters [5] and mode filter (in form of waveguide section with longitudinal slots) in order to measure transmission and reflection coefficients (Fig. 8). The measured transmission through the described uniform helical structure is shown in Fig. 9. The blue curve shows transmission through converters and mode filter (if we remove the structure at all). In this figure the calculated transmission coefficient is also shown so that one conclude there is 150 MHz shift of calculated and the obtained results. Instead of 29.93 GHz the structure is well matched at ~29.8 GHz. The reason is a small mistake in average radius of the structure. In case of the structure with Chebyshev's matchers there is broad band of good matching (Fig. 10).

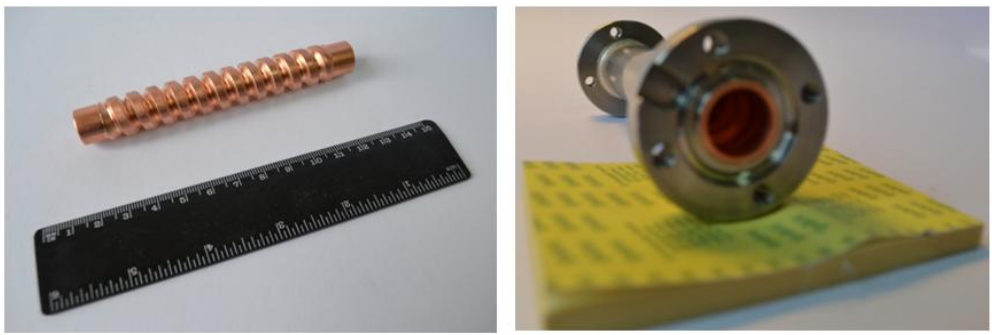

Figure 5: Helical structure photographs: main body (left) and helical structure supplemented by flanges (right).

Insert PSN Here

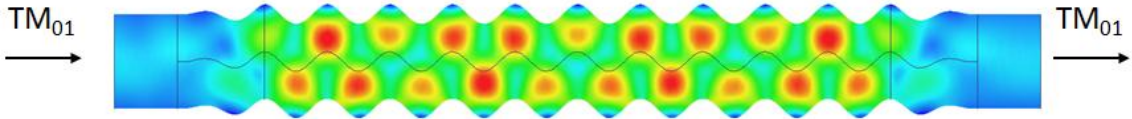

Figure 6: Helical structure matched with incoming TM01 mode in circular cross-section waveguide.

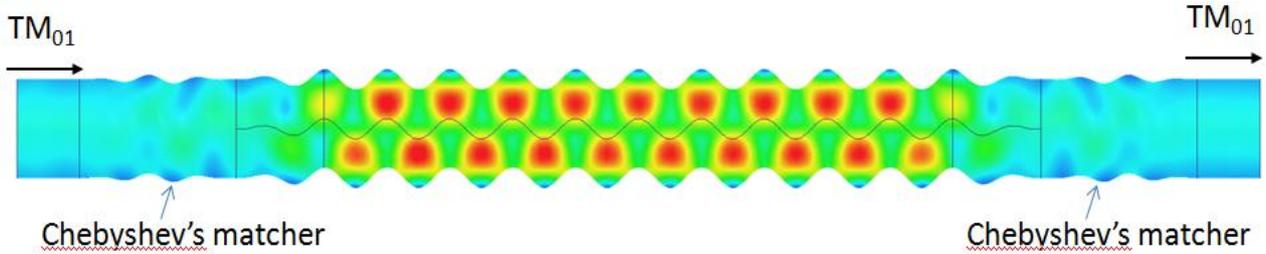

Figure 7: Helical structure matched by means of Chebyshev's matchers.

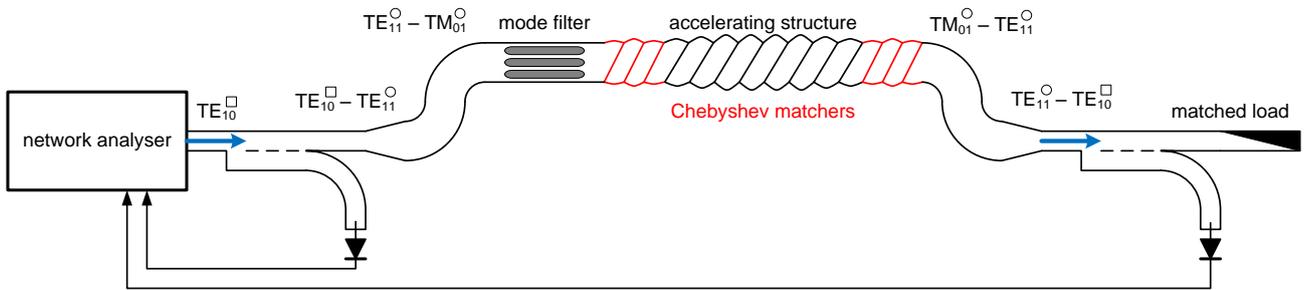

Figure 8: Scheme of low-power measurements.

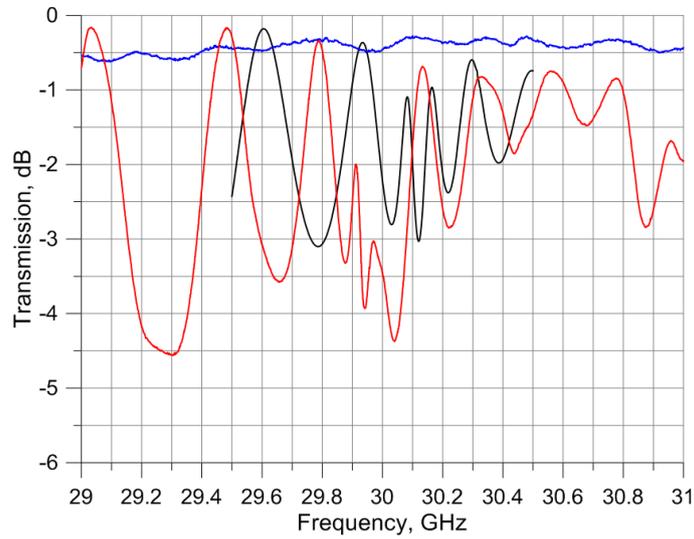

Figure 9: Transmission parameter of TM01-TM11 structure: measured (red curve), calculated (black curve) and reference level (blue curve).

Insert PSN Here

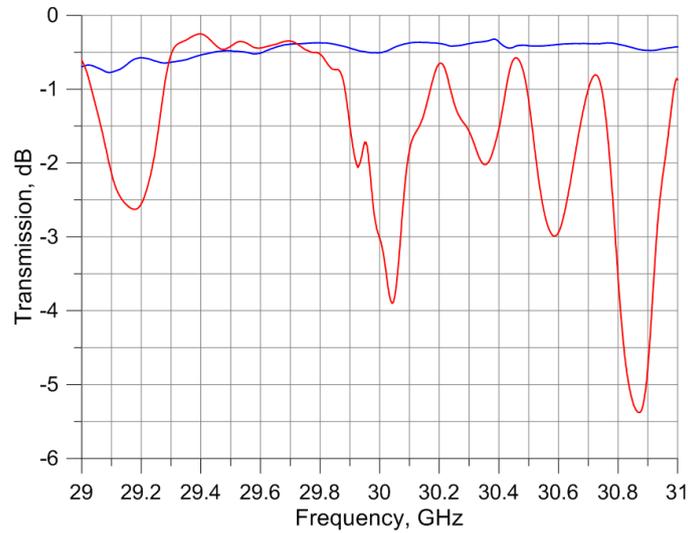

Figure 10: Tranmission parameter of TM01-TM11structure with added Chebyshev's matchers: measured (red curve), and reference level (blue curve).

## 3. SHUNT IMPEDANCE ENHANCEMENT

The described in the previous paragraphs helical structure has less shunt impedance in comparison with classical axisymmetrical accelerating structures at th same frequency. The shunt impedance can be enhanced in a structure which looks like "classical" one. Nevertheless, the idea of emittance preservation due to radiation cooling still works in this case. This new bi-periodic structure consists of periodic irises in circular cross-section waveguide which are slightly off-axis and periodically shifted each to other (Fig. 11). The selected parameters are: radius 4 mm, distance between irises 8.4 mm, periodicity of irises 16.8 mm, shift of irises 0.6 mm, width of irises 1 mm, and the hole diameters is 2.8 mm. The structure with these parameters provides as high shunt impedance as more than 90 MOhm/m. The dispersion for this structure is shown in Fig. 12. Of course, such structure has lost some appealing properties of the primary idea realized in the described helical structure.

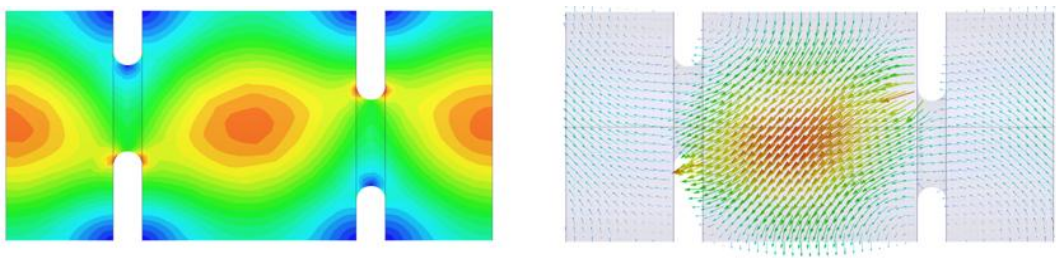

Figure 11: Operating mode of bi-periodic accelerating structure: complex electric field (left) and instantaneous electric field (right).

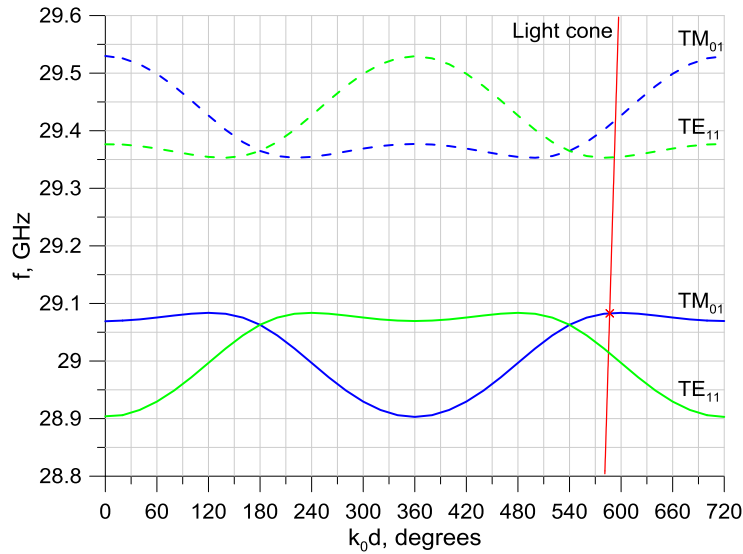

Figure 12: Frequency dispersion in bi-periodic structure.

## 4. CONCLUSION

The $TM_{01} - TM_{11}$ helical accelerating structure has several appealing properties. In particular, non-synchronous electric and magnetic field components are used, in order to preserve low beam emittance and small energy spread. The structure allows high accelerating gradient due to nanosecond filling time. The shunt impedance is slightly less than in a conventional accelerating structure. In order to increase shunt impedance, one might use the design based on irises with periodic off-axis shift. The carried out first low-power tests show promising results. High-power experiment with the use of multi-megawatt 30 GHz FEM to feed the structure is coming.

## Ackowledgments

Work supported by Russian Science Foundation grant № 14-19-01723.

Insert PSN Here